\documentclass[prb,twocolumn,showpacs,
preprintnumbers,amsmath,amssymb]{revtex4}
\usepackage{graphicx}% Include figure files
\usepackage{dcolumn}% Align table columns on decimal point
\usepackage{bm}% bold math

\begin{document}

%\preprint{APS/123-QED}

\title{Superconductivity on a M\"obius strip: numerical \\
studies on order parameter and quasiparticles}
% Force line breaks with \\

\author{Masahiko Hayashi}
 \email{hayashi@cmt.is.tohoku.ac.jp}
 %Lines break automatically or can be forced with \\
\author{Hiromichi Ebisawa}%
\affiliation{%
Graduate School of 
Information Sciences, Tohoku University, \\
Aramaki Aoba-ku, Sendai 980-8579, Japan\\
and JST-CREST
}%
\author{Kazuhiro Kuboki}
\affiliation{Department of Physics, Kobe University, Kobe 657-8501, Japan}

\date{\today}% It is always \today, today,
             %  but any date may be explicitly specified

\begin{abstract}
Superconducting states of an anisortopic 
s-wave superconductor on a M\"obius 
strip are studied numerically 
based on the Ginzburg-Landau theory 
and the Bogoliubov-de Gennes theory. 
In both, the equations are solved numerically on discitized lattice 
and the nonlinearity and the self-consistency are fully 
taken into account. 
First, we study the superconducting states on the M\"obius
strip in the presence of the Aharonov-Bohm flux threading the 
ring by employing the Ginzburg-Landau theory, and confirm the phase diagram 
previously proposed by Hayashi and Ebisawa 
[J. Phys. Soc. Jpn. {\bf 70}, 3495 (2002)]. 
The metastable states as well as the equilibrium energy state are 
studied and the nonequiriblium processes when the magnetic field 
is varied at a fixed temperature are discussed.  
Next, we study the microscopic superconducting states on the 
M\"obius strip based on the 
Bogoliubov-de Gennes theory, 
especially focusing on the state with a real-space node in 
the superconducting gap, which is expected to appear when the 
flux threading the ring is half the superconducting flux quantum.  
The local density of states in this {\it nodal state} is calculated in detail and the 
existence of the zero-energy bound states is shown. 
\end{abstract}

% PACS, the Physics and Astronomy
                             % Classification Scheme.
%\keywords{Suggested keywords}%Use showkeys class option if keyword
                              %display desired
\maketitle

\section{Introduction}

The realization of crystals with unusual
shapes, {\it e.g.}, ring, cylinder etc., by Tanda {\it et al.}
\cite{Tanda1,Tanda2,Okajima} has stimulated renewed interest
in the effects of the system geometry on the 
physical properties. 
Especially, the synthesis of M\"obius strip made of
transition metal calcogenides (NbSe$_3$, TaS$_3$ etc.)
\cite{Tanda}
opens new possibility to examine the physical properties of
superconductivity or charge density wave 
in topologically nontrivial spaces.

Recently several groups have studied
physical systems on M\"obius strips.
Hayashi and Ebisawa \cite{Hayashi-Ebisawa}
studied s-wave superconducting (SC) states on a M\"obius strip
based on the Ginzburg-Landau (GL) theory and found that
the Little-Parks oscillation, which is characteristic to the 
ring-shaped superconductor, is modified for the M\"obius strip
and a new state, which does not appear for
ordinary ring, shows up when the number of the magnetic flux quanta
threading the ring is close to a half odd integer. 
Yakubo, Avishai and Cohen \cite{Yakubo-Avishai-Cohen}
have studied the spectral properties
of the metallic M\"obius strip with impurities and 
clarified statistical characteristics of the fluctuation of 
the persistent current as a function of the magnetix flux threding the ring. 
The persistent current in a more simplified version of the M\"obius strip has also 
been studied by Mila, Stafford and Caponi \cite{Mila}. 
Wakabayashi and Harigaya \cite{Wakabayashi-Harigaya}
have studeid the M\"obius strip made of a nanographite ribbon, and 
the effects of M\"obius geometry on the edge localized states, which 
is peculiar to the graphite ribbon, has been clarified. 
A study from a more fundamental point of view can be found in the paper by 
Kaneda and Okabe \cite{Kaneda-Okabe}
where the Ising model on M\"obius strip 
and its domain wall structures are studied.

The main result of Ref. \onlinecite{Hayashi-Ebisawa}
is that if the magnetic flux threading the ring is close to a half 
odd integer times the 
flux quantum $\phi_0=hc/(2e)$ ($h$, $c$, $e$ being 
the Planck constant, the speed of right and the electron charge, respectively),  
a novel SC state appears. 
This state has a real-space node in SC gap 
along the strip: 
namely, the gap tends to zero along the line located 
in the middle of the strip. 
Throughout this paper we call this state the \lq\lq {\it nodal state}\rq\rq. 
It has been shown that the free energy of this state 
can be lower than that of uniformly gapped state, 
which is known to be the most 
stable state in case of the ordinary SC rings. 

This paper extends the previous study in the following two points: 
\begin{enumerate}
\item[a)] In Ref. \onlinecite{Hayashi-Ebisawa}, 
we have shown that the free energy of the {\it nodal state} is 
lower than other likely states. 
However, there is no evidence that it is the most stable. 
Although it is difficult to examine 
all possible local-minimum states of the nonlinear GL free 
energy, in this paper we try to give a more convincing 
evidence by resorting to a numerical method. 
We perform numerical minimization 
of the GL free energy and 
find as many local-minimum states as possible 
and reexamined the phase diagram of Ref. \onlinecite{Hayashi-Ebisawa}. 
A similar method is previously employed, 
for example, in Refs. \onlinecite{Schweigert} and 
\onlinecite{Baelus} in studying SC disks etc. 
\item[b)] The electronic states, which are not treated in Ref. \onlinecite{Hayashi-Ebisawa}, 
are studied in terms of the Bogoliubov-de Gennes (BdG) theory. 
We numerically solve the BdG equation on a lattice selfconsistently, 
where electrons are treated by the tight-binding approximation. 
Based on the solution we study the local density of states 
in the {\it nodal state}, which may be observed, for example, by 
scanning tunneling microscope measurement. 
\end{enumerate}

This paper is organized as follows: 
In Sec. \ref{review}, 
we review the behavior of a SC M\"obius strip in a magnetic field, 
presented in Ref. \onlinecite{Hayashi-Ebisawa}. 
In Sec. \ref{gl}, the model, method and results of 
the numerical analyses of GL theory are presented. 
In Sec. \ref{bdg}, studies based on the BdG theory for the electronic properties of 
the {\it nodal state} are presented. 
In Sec. \ref{discussion}, discussions on the results and their relation to 
experiments are given. 
Sec. \ref{summary} is devoted to summary. 

\section{A superconducting M\"obius strip in a magnetic field}
\label{review}

%%%%%{FIG}%%%%%%
\begin{figure}[t]
\begin{center}
\includegraphics[width=7cm,clip]{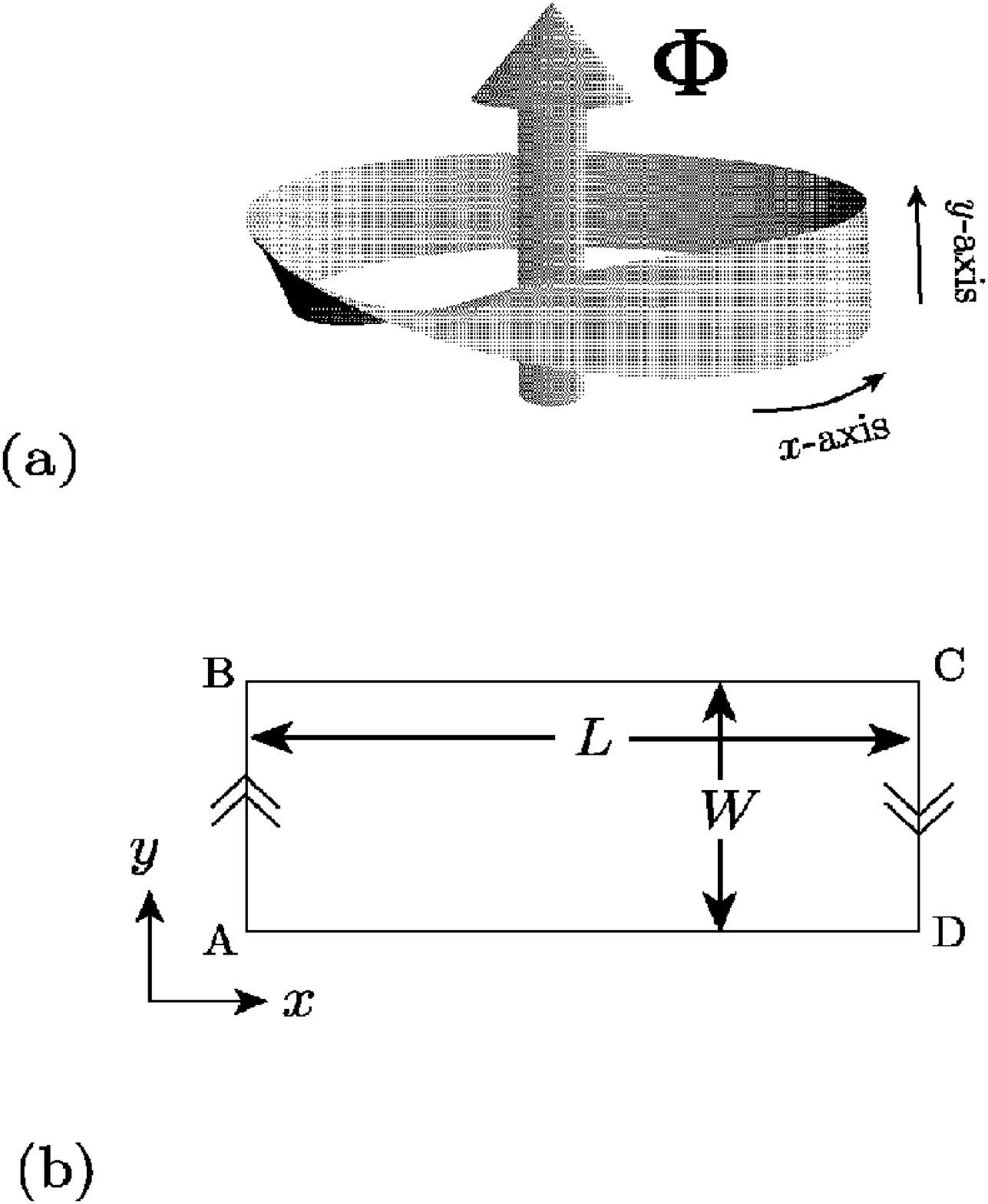}
\caption{
(a) Geometry of SC M\"obius strip 
where an Aharonov-Bohm flux is threading the 
ring (b) The developed figure of (a).  }
\label{strip}
\end{center}
\end{figure}
%%%%%{FIG}%%%%%%
%%%%%{FIG}%%%%%%
\begin{figure}[t]
\begin{center}
\includegraphics[width=7cm,clip]{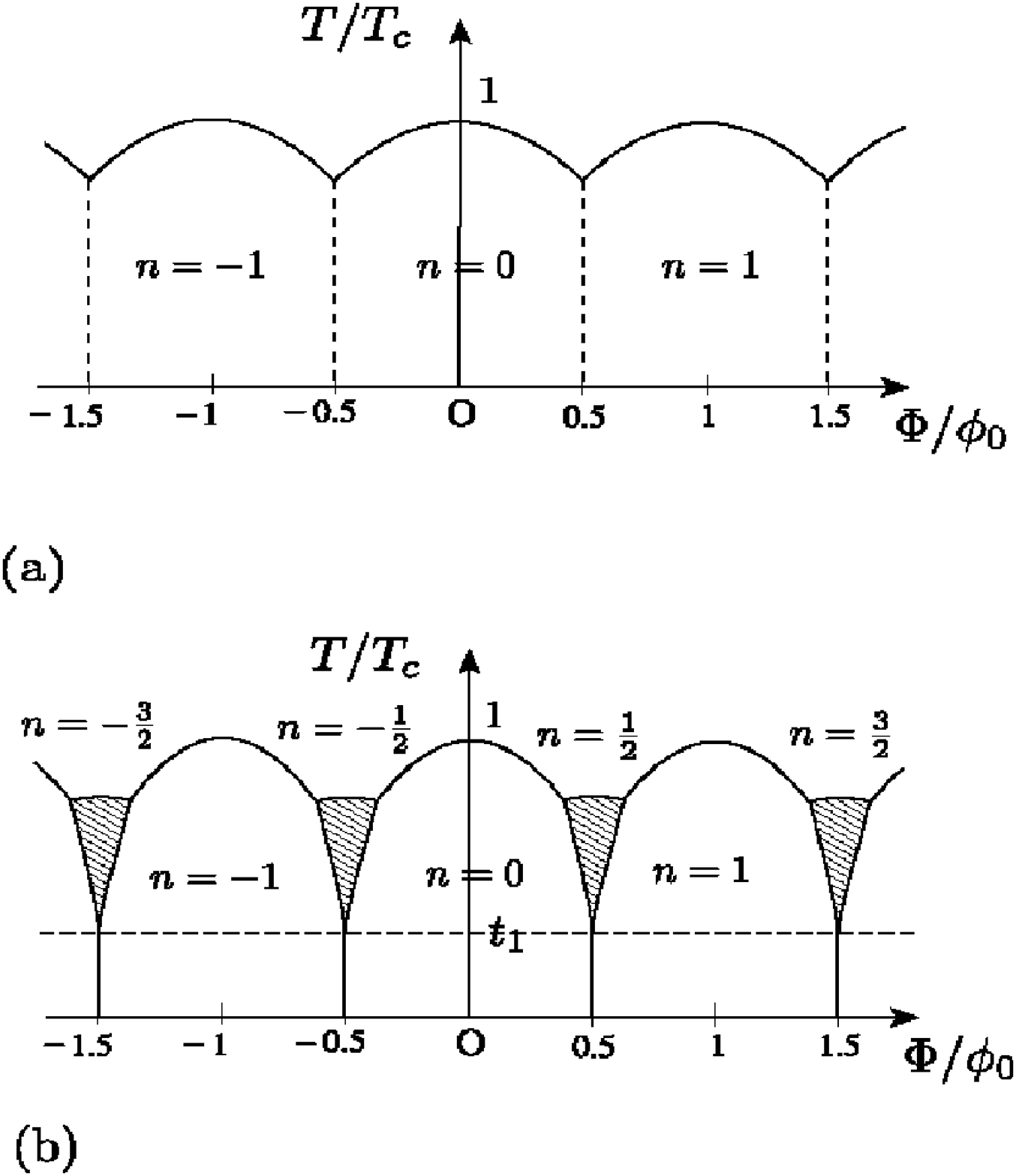}
\caption{
Phase diagram of a sperconducting M\"obius 
strip in the presence of a AB flux with 
(a) $\frac{\pi r_\parallel}{2\sqrt{3}} < r_\perp$ and  
(b) $\frac{\pi r_\parallel}{2\sqrt{3}} > r_\perp$. }
\label{free2}
\end{center}
\end{figure}
%%%%%{FIG}%%%%%%

Here we summarize the results obtained in Ref. \onlinecite{Hayashi-Ebisawa}. 
In that paper, the behavior of a SC M\"obius strip in an 
external magnetic field is studied based on the GL theory. 

We consider the M\"obius strip made of a superconductor
as shown in Fig. \ref{strip} (a). 
The magnetic field is assumed to be threading the ring 
in a form of the so-called Aharonov-Bohm (AB) flux
as indicated in the figure by a bold arrow, 
which gives rise to a non-zero vector potential 
on the strip although the magnetic field is vanishing on the strip. 
We further assume that 
the coherence length along the strip and that perpendicular 
to the strip, which are respectively denoted as $\xi_{\parallel}$ 
and $\xi_{\perp}$, are different. 
Throughout this paper we set $x$-axis and $y$-axis as 
indicated in the figure. 
The developed figure of the M\"obius strip is given 
in Fig. \ref{strip} (b). 
We denote the circumference and the width 
by $L$ and $W$, respectively. 

%%%%%{FIG}%%%%%%
\begin{figure}[th]
\begin{center}
\includegraphics[width=6cm,clip]{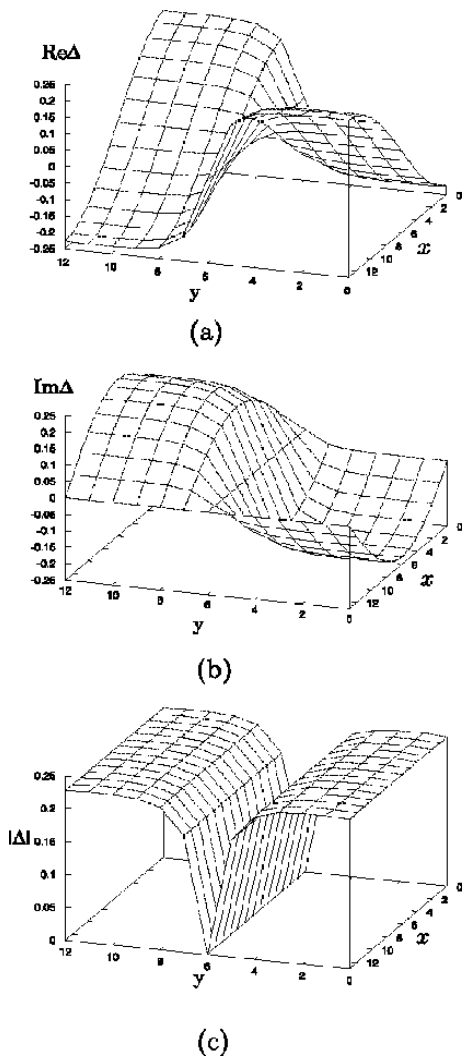}
\caption{(a) Real part, 
(b) imaginray part and (c) amplitude of the order parameter 
in the {\it nodal state}. 
%This figures are obtained by solving 
%the BdG equation, as stated in Chap. \ref{bdg}. 
}
\label{moebius_op}
\end{center}
\end{figure}
%%%%%{FIG}%%%%%%

In this paper the effects of the bending, or the 
non-zero curvature, caused in the strip by the 
M\"obius geometry, are neglected. 
Although these may be important in the case of unconventional 
pairing, such as $p$-wave superconductors \cite{Sigrist-Ueda}, 
they may be neglected for 
$s$-wave superconductors, treated in this paper.

The equiriblium states of SC M\"obius strip depends on the 
magnetic flux $\Phi$ threading the ring. 
The SC M\"obius strip 
shows the so-called Little-Parks oscillation of the 
transition temperature with a period $\phi_0$, 
as we naively expect from 
the analogy to the ordinary rings.  
However, it turned out that the oscillation can be appreciably 
modified in SC M\"obius strip 
depending on the strength of the anisotropy 
of the coherence lengths. 

The analysis of Ref. \onlinecite{Hayashi-Ebisawa} shows that 
there are two important parameters, which are given by 
\begin{align}
r_{\perp}=\frac{\xi_{\perp}(0)}{W},\,\,\,\,\,\,
r_{\parallel}=\frac{\xi_{\parallel}(0)}{L}, 
\end{align}
where $\xi_{\perp}(0)$ and $\xi_{\parallel}(0)$ are the  
coherence lengths at absolute zero temperature($T=0$). 

It has been shown that 
when the condition $\frac{\pi}{2 \sqrt{3}} r_{\parallel} < r_{\perp}$
is satisfied, the phase diagram in a magnetic field 
behaves like the one shown in Fig. \ref{free2} (a), 
which is basically the same as in the case of ordinary ring. 
Here $T_c$ is the transition temperature in the bulk and 
the index $n$ indicated in the figure denotes the 
the winding number of the phase as we go around the 
ring once along a trajectory parallel to the edge. 
When 
\begin{align}
\frac{\pi}{2 \sqrt{3}} r_{\parallel} > r_{\perp}
\label{condition}
\end{align}
is satisfied, a state characteristic to the 
M\"obius geometry appears when the number of the flux is close to 
a half odd integer, as shown in Fig. \ref{free2} (b) by hatched regions. 
These states are indexed with a half odd $n$ and 
the phase of the order parameter changes by an odd number times 
$\pi$ as we go around ring once. 
The spatial dependence of the order parameter is shown in Fig. \ref{moebius_op}, 
where the real part, the imaginary part and the amplitude 
are shown for $n=1/2$ case. 
It is clear that the order parameter has a real-space node 
in the middle of the strip. 
Therefore we call these states the \lq\lq {\it nodal states}\rq\rq. 
The {\it nodal states} can exist above $t_1$ shown in Fig. \ref{free2} (b). 
Here $t_1$ is given by 
\begin{align}
t_1=1-\left(\frac{3 \pi^2}{4 \sqrt{2}}\frac{r_{\parallel}^2}{r_{\perp}}
\right)^2
\end{align}
as one can see from the results in Ref. \onlinecite{Hayashi-Ebisawa}. 

The results of Ref. \onlinecite{Hayashi-Ebisawa} are 
obtained by comparing the free energies of the several possible states, 
which are chosen empirically. 
Therefore it is not easy to say 
that there is no states with lower free energies. 
Of course, it is impossible to investigate all 
possible order parameter configurations. 
However more reliable analysis, which 
can cover wider range of the configuration space is 
required. 
In this paper, to fulfill this requirement, we perform numerical 
study on GL theory, 
which is given in Sec. \ref{gl}. 

The analysis of Ref. \onlinecite{Hayashi-Ebisawa} is 
limited to the phenomenological one. 
Since the {\it nodal state} is also interesting from electronic 
points of view, more microscopic study is required. 
We solve BdG equations on the 
M\"obius strip numerically and clarify the electronic bound 
state near the node of the {\it nodal state}. 
This will be given in Sec. \ref{bdg}.

\section{Numericl study based on Ginzburg-Landau thoery}
\label{gl}

In this section, we study the SC state on the M\"obius strip numerically 
using the GL theory. 
Here we employ non-linear optimization method (quasi-Newton method) 
to find the local minimum state of the GL free energy \cite{recipi}. 
%In this method, first we set an initial value for the order parameter and, 
%then, vary it until we finally reach the local minimum. 
%Therefore the obtained value depends on the choice of the initial state. 
%In order to find as many local minimum states as possible, we tried many 
%randomly generated initial states. 
%These procedure is same as one adopted in Refs. \onlinecite{Schweigert,Baelus}. 

GL free energy $F$ of our system is given as follows:
\begin{align}
F & =d_\perp\int d^2\vec{r}\Biggl[\frac{\hbar^2}{2 m^*}\biggl\{
\left|\left(\frac{\partial}{\partial x}-i\frac{2 \pi}{\phi_0} A_x\right)\psi\right|^2\nonumber \\
& \phantom{=\int_V d\vec{r}\biggl[}+\gamma^2
\left|\left(\frac{\partial}{\partial y}-i\frac{2 \pi}{\phi_0} A_y\right)
\psi\right|^2\biggr\}\nonumber \\
 & \phantom{=\int_V d\vec{r}\biggl[}+\alpha_0\left(t-1\right)|\psi|^2 +\frac{\beta}{2}|\psi|^4\Biggr].
\end{align}
Here $m^*$, $\phi_0=hc/(2e)$ are the mass of a Cooper pair (twice the electron mass) 
and the magnetic flux quantum, respectively. 
The thickness of the strip $d_\perp$ 
is much smaller than the superconducting coherence length and 
the strip can be treated as two-dimensional. 
$\psi(\vec{r})$ and $\vec{A}(\vec{r})$ are the SC order parameter and the vector potential, respectively. 
$t=T/T_c$ is the reduced temperature. 
The AB flux is incorporated by taking 
$\vec{A} =\Phi/L \, \vec{e}_x $, 
where $\vec{e}_x$ is the unit vector in $x$-direction.
$\alpha_0$ and $\beta$ are positive constants, and 
$\gamma=\xi_\perp/\xi_\parallel$ is the anisotropy parameter.

For numerical calculations we introduce 
the lattice version of $F$ as 
\begin{align}
F=&F_0\biggl[\sum_{j=1}^{N_x}\sum_{k=1}^{N_y}\tilde{\xi}^2\biggl\{
|\tilde{\psi}(j,k)|^2+|\tilde{\psi}(j+1,k)|^2
\nonumber\\
&\phantom{\tilde{\xi}^2\biggl\{}
-\tilde{\psi}^*(j,k) \tilde{\psi}(j+1,k)e^{-i \tilde{a}_x}-{\rm c.c.}\biggr\}
\nonumber\\
&
+\gamma^2 \sum_{j=1}^{N_x}\sum_{k=1}^{N_y-1}\tilde{\xi}^2
\left|\tilde{\psi}(j,k) - \tilde{\psi}(j,k+1)\right|^2
\nonumber\\
&+\sum_{i=1}^{N_x}\sum_{j=1}^{N_y}
\biggl\{
(t-1)|\tilde{\psi}(j,k)|^2+\frac{1}{2}|\tilde{\psi}(j,k)|^4\biggr\}
\biggr], \label{free_lat}
\end{align}
where
$\tilde{a}_x =2 \pi d A_x/\phi_0$, 
$\tilde{\xi}$ is given by $\xi_\parallel(0)/d$, 
where $\xi_\parallel(0)=\sqrt{\hbar^2/2 m^* \alpha_0}$  
and $d$ is the lattice spacing. 
Here the lattice is assumed to be a square one. 
Note that $N_x d =L$ and $N_y d=W$. 
$F_0$ is $V_0 \alpha^2/\beta$ where 
$V_0=d^2\times d_\perp$. 
The order parameter $\tilde{\psi}$ is normalized so that $\tilde{\psi}\rightarrow 1$
as $t\rightarrow 0$. 

We use the open boundary condition in 
$y$-direction and M\"obius boundary condition 
in $x$-direction: 
namely, we put
\begin{align}
\tilde{\psi}(N_x+1,k)=\tilde{\psi}(1,N_y+1-k), 
\end{align}
($1 \le k \le N_y$) in the first two lines of Eq. (\ref{free_lat}). 

The order parameter $\tilde{\psi}(j,k)$ minimizing $F$ 
is obtained by solving the equation, 
\begin{align}
\frac{\partial F}{\partial \tilde{\psi}^*(j,k)}=0,
\label{eqns}
\end{align}
which yields, 
\begin{align}
&\tilde{\xi}^2\biggl\{-\tilde{\psi}(j+1,k)e^{-i \tilde{a}_x}
-\tilde{\psi}(j-1,k)e^{i \tilde{a}_x}
\nonumber\\
&\phantom{\tilde{\xi}^2\biggl\{}
-\gamma^2\tilde{\psi}(j,k+1)-\gamma^2\tilde{\psi}(j,k-1)+ 2(1+\gamma^2) \tilde{\psi}(j,k)\biggr\}
\nonumber\\
&\phantom{\tilde{\xi}^2\biggl\{}
+(t-1) \tilde{\psi}(j,k)+|\tilde{\psi}(j,k)|^2\tilde{\psi}(j,k)=0.
\label{gl1}
\end{align}

We obtained the solution of the Eq. (\ref{gl1}) 
in terms of the nonlinear optimization of the free energy Eq. (\ref{free_lat}). 
In this paper we especially utilized the so-called quasi-Newton method 
\cite{recipi}. 
Like other methods of nonlinear optimization, 
quasi-Newton method starts from an initial value of $\tilde{\psi}$ 
and changes it so that the free energy becomes lower  
until finally we reach the local minimum. 
Therefore which local minimum we reach depends on the 
initial value of $\tilde{\psi}$. 
In this paper we randomly chose the initial values and 
performed optimization as many times as possible. 
Then we obtained several local-minimum states. 
These procedure is the same as the one adopted 
in Refs. \onlinecite{Schweigert} and \onlinecite{Baelus}. 

The results are as follows. 
The system size we used is $N_x=10$, $N_y=10$, 
$\xi_0=1.5 d$ and $\xi_\perp=1.2 d$. 

%%%%%{FIG}%%%%%%
\begin{figure}
\begin{center}
\includegraphics[width=8cm,clip]{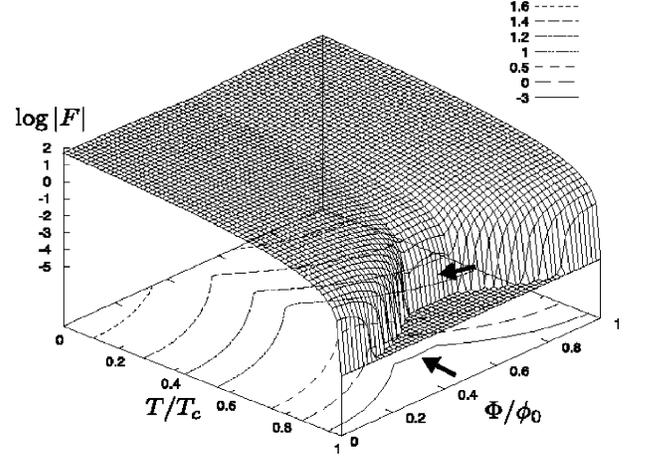}
\caption{
The free energy of the most stable state as a function of 
$T$ and $\Phi$. 
Around $\Phi\simeq \phi_0/2$ a structure corresponding to 
the {\it nodal state} can be seen. 
The vertical axis is the logarithm of the $|F|$ and 
the region with $|F|<10^{-5}$ near $T_c$ has been cut off. 
The free energy is measured in the unit of $F_0$. 
}
\label{phase_diag}
\end{center}
\end{figure}
%%%%%{FIG}%%%%%%

%%%%%{FIG}%%%%%%
\begin{figure}
\begin{center}
\includegraphics[width=8cm,clip]{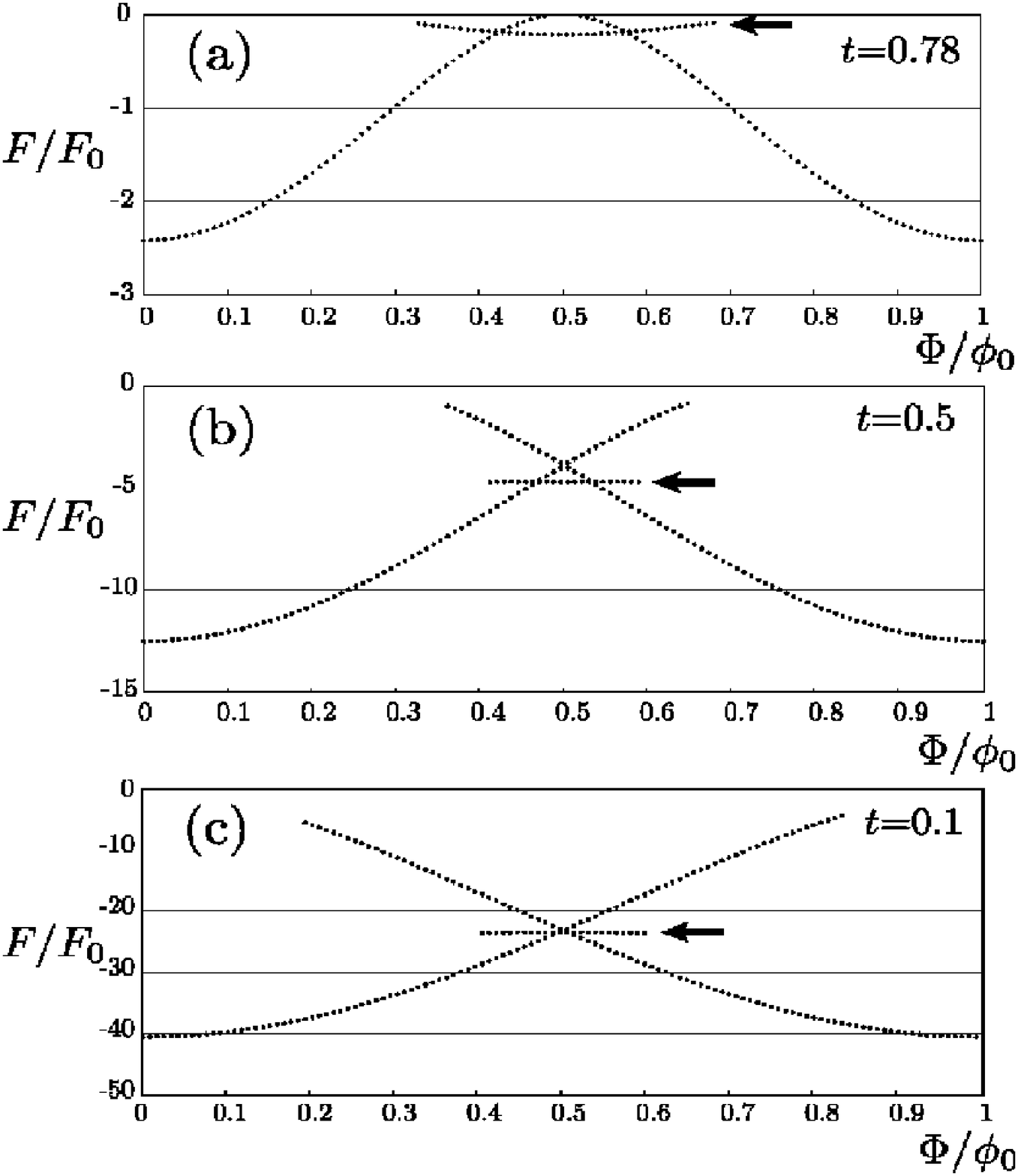}
\caption{
The free energy of the metastable states as a function of 
$\Phi$ for (a) $t=0.78$, (b) $t=0.5$, and (c) $t=0.1$. 
The branchs corresponding to the {\it nodal states} are 
indicated by bold arrows in every graph. 
It becomes the most stable state when $t=0.78$ and $0.5$ 
for a certain region of $\Phi$. 
When $t=0.1$ there is no stable region for the {\it nodal state}. 
}
\label{free}
\end{center}
\end{figure}
%%%%%{FIG}%%%%%%

First we discuss the phase diagram. 
The numerically obtained 
free energy is shown in Fig. \ref{phase_diag} 
using the log scale: 
we have shown only the region with $F <0$ and 
$\log |F|$ is plotted as a function of $\Phi$ and $T$. 
We can see from this figure that the 
structure corresponding to the {\it nodal state} 
appears near $\Phi=\phi_0/2$ 
(shown by bold arrows). 
Actually, the order parameter in this region behaves 
like that shown in Fig. \ref{moebius_op}. 

In Fig. \ref{free}, we have depicted the free energies of 
the metastable states at three different temperatures 
((a) $t=0.78$, (b) $t=0.5$ and (c) $t=0.1$). 
We clearly see three distinct series of states in each graph. 
The branch which starts from $\phi=0$ corresponds to 
the state with a uniform order parameter $\Delta={\rm (const.)}$ 
and that ends at $\phi=\phi_0$ corresponds to the state with 
$\Delta={\rm (const.)}\times e^{i 2 \pi x/L}$. 
Between these two branches, a branch corresponding to the 
{\it nodal state} can be seen. 
At $t=0.78$ and $t=0.5$, 
the {\it nodal state} has the lowest free energy near 
$\phi\simeq \phi_0/2$. 
However, at $t=0.1$, although the {\it nodal state} is still a metastable 
state, there is no region where the {\it nodal state} is the most stable. 
These features agrees with the prediction of 
Ref. \onlinecite{Hayashi-Ebisawa}.

\section{Numericl study based on Bogoliubov-de Gennes thoery}
\label{bdg}

In order to study the $s$-wave SC state on a
M\"obius strip microscopically, 
we treat a tight-binding model on a square lattice with attractive
on-site interactions.
The Hamiltonian of the system is given as
\begin{equation}
\begin{array}{rl}
   H =&  - \displaystyle t_x \sum_{{j}\sigma}
     (e^{i\phi_x}c_{{j}+{\hat x},\sigma}^{\dagger}c_{{j},\sigma}
     + e^{-i\phi_x}c_{{j},\sigma}^{\dagger}c_{{j}+{\hat x},\sigma})  \\
      & - \displaystyle t_y \sum_{{j}\sigma} (
      c_{{j}+{\hat y},\sigma}^{\dagger}c_{{j},\sigma} + 
      c_{{j},\sigma}^{\dagger}c_{{j}+{\hat y},\sigma}) \\
      & -  \displaystyle V \sum_{j} 
      n_{{j}\uparrow}n_{{j}\downarrow}
      - \mu \sum_{{j}\sigma} 
      c_{{j},\sigma}^{\dagger} c_{{j},\sigma}
\end{array}
\end{equation}
where $c_{j\sigma}$ is the annihilation operator of electron 
at a site $j$ with spin $\sigma (= \uparrow,\downarrow)$, 
and $V (>0)$ and $\mu$ are the attractive interaction and the chemical
potential, respectively. 
Here ${j}=(j_x,j_y)$ ($1 \leq j_x \leq N_x, 1\leq j_y\leq N_y$) 
numbers the sites, 
where $N_x$ and $N_y$ are the numbers of sites
along the $x$- and $y$-direction, respectively, and 
${\hat x}=(1,0)$, ${\hat y}=(0,1)$. 
$n_{{j} \sigma} = c^{\dagger}_{{j} \sigma} c_{{j} \sigma}$ is the 
electron number operator. 
The transfer integrals for $x$- and $y$-directions are denoted as $t_x$ and $t_y$, 
respectively, 
and the Peierls phase $\phi_x = (\pi/N_x)(\Phi/\phi_0)$
represents the effect of the AB flux $\Phi$ threading the M\"obius strip. 
To realize the M\"obius geometry we impose, for $x$-direction, the condition, 
\begin{equation}
\displaystyle c_{(N_x+1,j_y),\sigma} = c_{(1,N_y-j_y+1),\sigma},
\end{equation}
while for the $y$-direction we assume the open boundary condition.  

The interaction term is decoupled within a mean-field approximation as 
\begin{equation}
n_{{j}\uparrow}n_{{j}\downarrow} \to
\Delta_{{j}}c_{{j}\downarrow}^\dagger c_{{j}\uparrow}^\dagger
+\Delta_{{j}}^{*}c_{{j}\uparrow} c_{{j}\downarrow}
-\vert \Delta_{{j}} \vert^2
\end{equation}
with $\Delta_{{j}} \equiv \langle c_{{j}\uparrow}c_{{j}\downarrow}\rangle$
being the SC order parameter.
Then the mean-field Hamiltonian is written as
\begin{equation}
{\cal H}_{\rm MFA} = \sum_{j}\sum_{k} \Psi_{j}^\dagger 
{h}_{{j}{k}} \Psi_{k}
\end{equation}
where
\begin{equation}
{h}_{{j}{k}}=
\left [\begin{array}{cc}
W_{{j}{k}} & F_{{j}{k}} \\
F_{{j}{k}}^{*} & -W_{{j}{k}}
\end{array}\right ], \ \
%\end{equation}
% \begin{equation}
\displaystyle \Psi_j \equiv 
\left[
\begin{array}{l}
c_{{j}\uparrow}^\dagger \\
c_{{j}\downarrow}
\end{array}
\right] .
\end{equation}
with
\begin{equation}
\begin{array}{rl}
W_{{j}{k}}= &\displaystyle
-t_x(e^{i \phi_x}\delta_{{k},{j}+{\hat x}}+
e^{-i \phi_x} \delta_{{k},{j}-{\hat x}})\\
&-t_y (
\delta_{{k},{j}+{\hat y}}+
\delta_{{k},{j}-{\hat y}})
- \mu\delta_{{j}{k}}, \\
F_{{j}{k}} = & - \Delta_{{j}} \delta_{{j}{k}}.
\end{array}
\end{equation}
By solving the following BdG equation
\begin{equation} 
\sum_l h_{jl} 
\left (\begin{array}{cc}
u_{ln}   \\
v_{ln}
\end{array}\right )
= E_n
\left (\begin{array}{cc}
u_{jn}   \\
v_{jn}
\end{array}\right )  
\end{equation} 
we can obtain the energy eigenvalues $E_n$ and the
corresponding eigenfunctions $(u_{jn}, v_{jn})$, where $n$ is 
numbering the states.
The unitary transformation using $(u_{jn}, v_{jn})$ 
diagonalizes ${\cal H}_{\rm MFA}$, and
the SC order parameter $\Delta_{{j}}$ 
can be written in terms of $E_n$ and $(u_{jn}, v_{jn})$.
These constitute the self-consistency equations which
will be solved numerically. 
In the following we take the parameters as,
\begin{align}
&t_x=1.0,\,\,\,t_y=0.49\nonumber\\
& \mu=0,\,\,\,T=0.22,\,\,\,V=0.25. 
\label{param}
\end{align}

%%%%%{FIG}%%%%%%
\begin{figure}[htb]
\begin{center}
\includegraphics[width=7.5cm,clip]{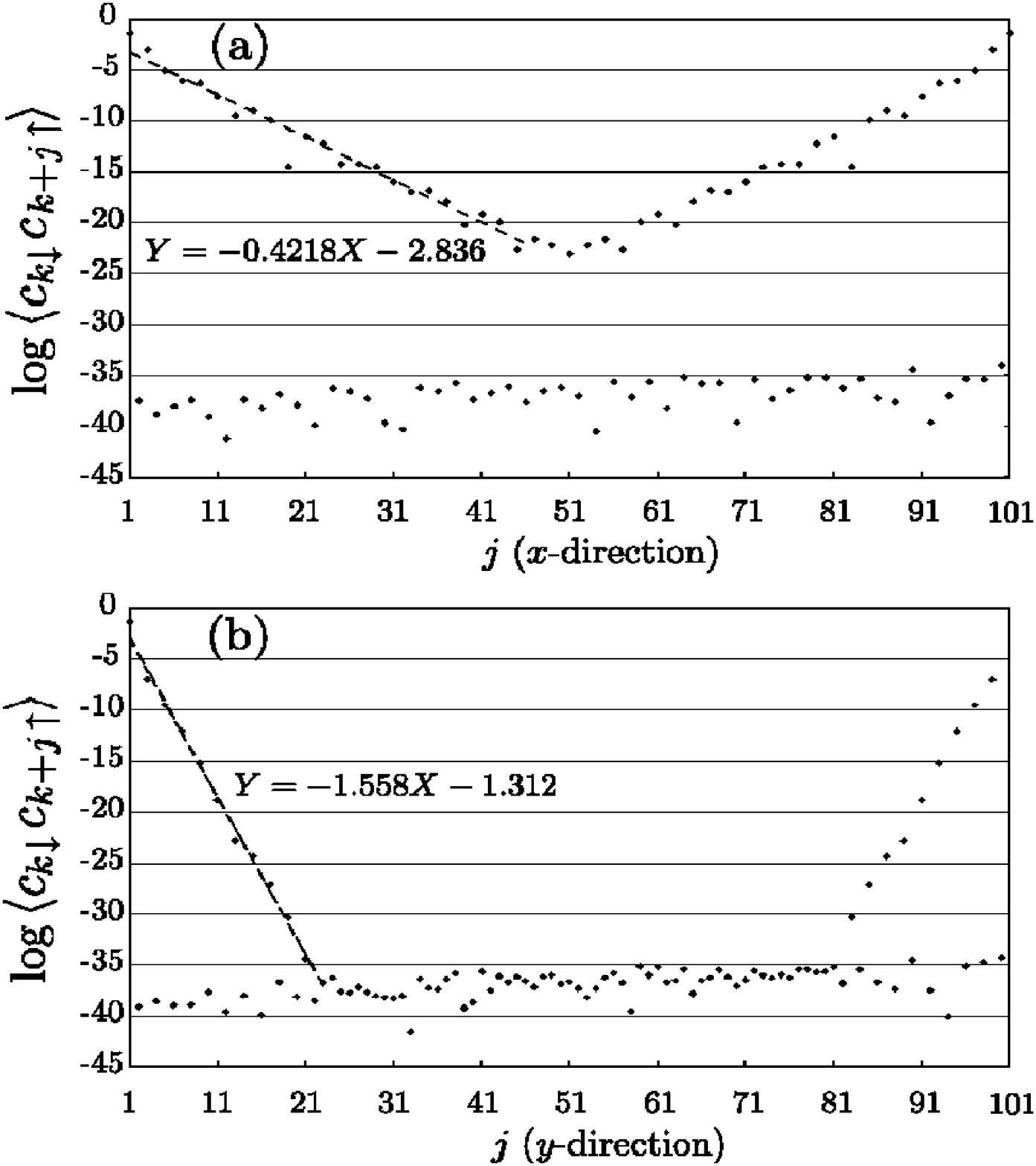}
\caption{
The anomalous correlation function 
$\langle c_{k \downarrow} c_{k+j \uparrow}\rangle$
for the system with the parameters in Eq. (\ref{param}). 
(a) and (b) correspond to the correlation in 
the $x$- and $y$-direction, respectively. 
The equations in the figures are the 
approximating lines to the 
exponential decay. }
\label{corr}
\end{center}
\end{figure}
%%%%%{FIG}%%%%%%

First we estimate the correlation lengths, $\xi_{\parallel}(0)$ and $\xi_{\perp}(0)$, 
by calculating the anomalous Green's function 
$F(j)\equiv\langle c_{k \downarrow} c_{k+j \uparrow}\rangle$ 
at $T=0$ without applying the magnetic flux, 
and the results are depicted 
in Fig. \ref{corr}. 
(This calculation has been carried out in a larger system with 
$100 \times 100$ sites using 
periodic boundary condition for both $x$- and $y$-directions.)
From the results in Fig. \ref{corr} and the relation 
$F(j)\propto \exp(-|j|/\xi)$, 
$\xi_\parallel(0)$ and $\xi_\perp(0)$ 
are estimated as 
\begin{align}
\xi_\parallel(0) = 2.37 d,\,\,\,\xi_\perp(0) = 0.64 d, 
\end{align}
where $d$ is the lattice spacing. 

The calculation in the M\"obius geometry is carried out for 
a system with $N_x=13$ and $N_y=14$. 
Putting $L=13\, d$ and $W=14\, d$, 
we obtain $r_\parallel = 0.18$ and $r_\perp=0.046$. 
These values satisfies the condition, Eq. (\ref{condition}). 
Furthermore, $t_1$ is estimated to be $-13$, 
which means that the {\it nodal state} is stable at all 
temperature range down to 0K. 
In passing, the bulk $T_c$ of this system is $0.358$ 
(estimated numerically 
in the system with $100\times 100$ sites), 
which coincides with that in $13\times 14$ system with $\Phi=0$ 
within a numerical accuracy. 

\begin{figure}[htb]
\begin{center}
\includegraphics[width=8cm,clip]{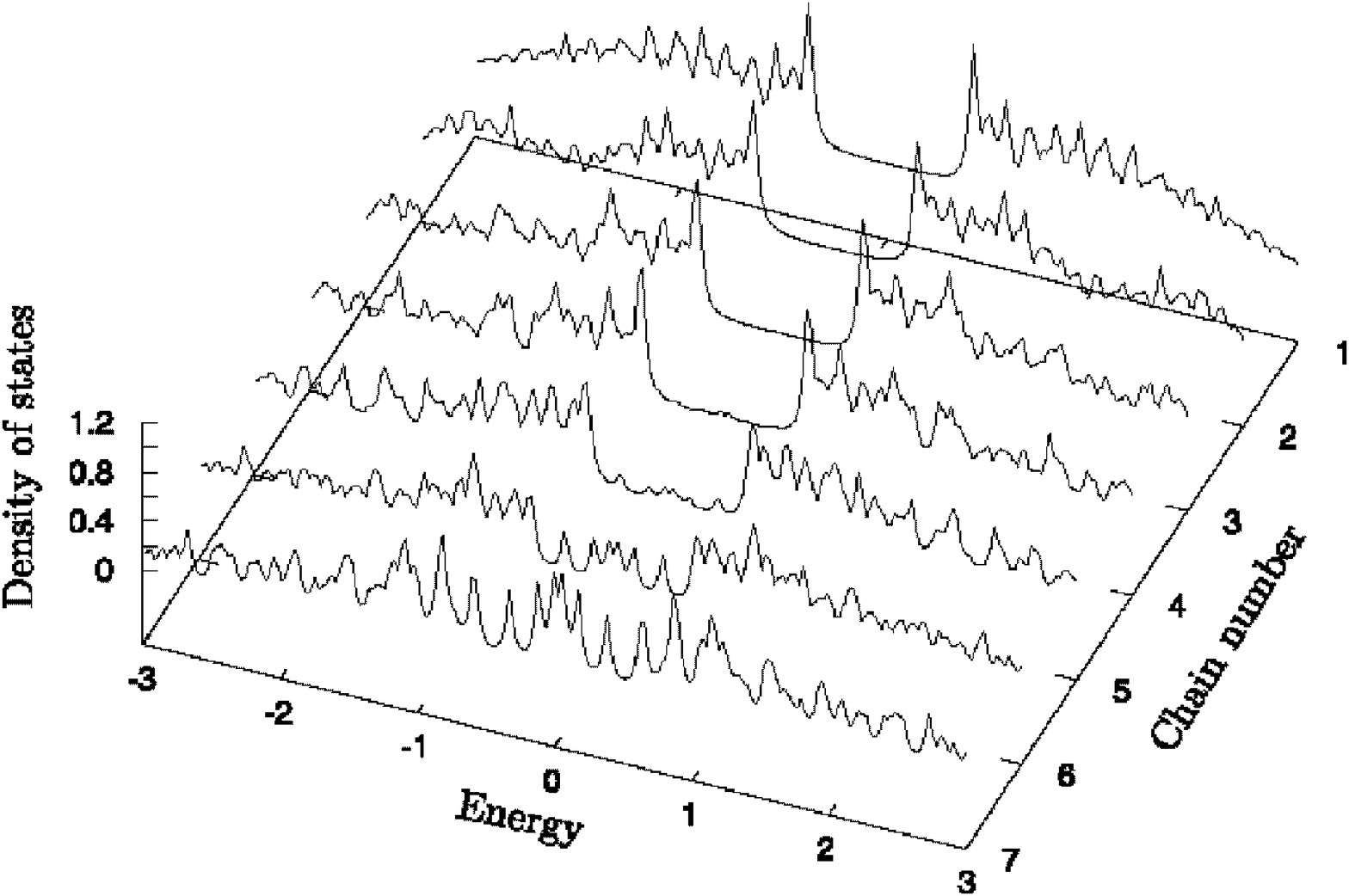}
\caption{
The local density of state in the {\it nodal state} on 
the M\"obius strip}
\label{dos}
\end{center}
\end{figure}

In this calculation, we limited ourselves to the case of 
the {\it nodal state} at $\phi=\phi_0/2$. 
Since this microscopic calculation needs more time to 
obtain good convergence as compared to GL calculation, 
we have selected for the initial order parameter the solution obtained by 
the GL analysis, such as the one depicted in Fig. \ref{moebius_op}. 
After the calculation, the behavior of the order parameter 
has not changed so much and we consider that the 
iteration converged to the {\it nodal state}. 
The local density of states (LDOS) is calculated 
from the equation, 
\begin{align}
N(j,E)= -\frac{1}{\pi} {\rm Im} 
\sum_n \frac{u_{jn}^* u_{jn}}{E-E_n+i \Gamma}. 
\end{align}
where $\Gamma$ is the broadening of the single energy level, 
introduced to simulate the actual experiment.     
The result is shown in Fig. \ref{dos} with 
$\Gamma=0.03$. 
Each line of Fig. \ref{dos} corresponds to the LDOS at 
$j=(1,j_y)$ where $j_y$ numbers the chain 
from the edge of the strip. 
(The LDOS is independent of $j_x$.)
Because of the inversion symmetry with respect to the center, 
we depicted LDOS only for $1 \le j_y \le7$, 
where $j_y=1$ and $j_y=7$ correspond to the outermost and the 
innermost chain, respectively. 

In Fig. \ref{dos}, a well-developed 
gap behavior can be seen for $1 \le j_y \le5$ and 
the bound states are formed in 
the chains $j_y= 6$ and $7$. 
These bound states originates from the node. 
Since the node can be regarded as the spontaneously 
formed $\pi$-junction (as one can see from 
Fig. \ref{moebius_op}, the phase of the order parameter 
changes by $\pi$ in crossing the node), the bound 
state energy must be zero. 
In Fig. \ref{dos}, the deviation from zero energy is 
seen, which may be due to the finite size effect. 
This point will be discussed in detail in a separate paper
\cite{Suzuki_et_al}. 

%%%%%{FIG}%%%%%%
%%%%%{FIG}%%%%%%

%%%%%{FIG}%%%%%%
\begin{figure}[htb]
\begin{center}
\includegraphics[width=6cm,clip]{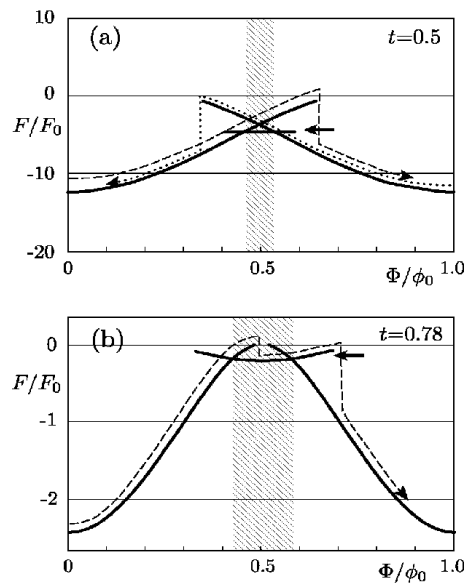}
\caption{
The transition of the states in the field sweep 
experiment for (a) $t=0.5$ and (b) $t=0.78$. 
The {\it nodal state} is the most stable in the shaded region. 
See text for the details. 
}
\label{sweep}
\end{center}
\end{figure}
%%%%%{FIG}%%%%%%

\section{Discussion}
\label{discussion}
In this paper, by using the numerical methods, we have
confirmed that the predictions of Ref. \onlinecite{Hayashi-Ebisawa}
are satisfied within both GL and BdG level. 
Until now, we have limited our discussion to 
purely two-dimensional cases. 
This is not the case for actually synthesized M\"obius crystals 
\cite{Tanda}. 
Here we discuss the effects of the finite thickness of the strip. 
If the M\"obius strip is thicker than the SC 
coherence length perpendicular to the strip surface, 
the {\it nodal state} may no longer be stable and 
the node, which is obtained under the assumption that 
the gap is uniform in the direction of the thickness, 
may become a vortex line, embedded inside the strip. 
Then there is no gapless region on the surface. 
In this case, we can not observe the node by only measuring 
the surface density of states by scanning tunneling microscope.  
The observation of the {\it nodal state} may be possible 
only by measuring the variation of the gap 
or magnetization as a function of the magnetic flux. 
More precise analysis on the thicker M\"obius strips 
is left for the future study. 

In Sec. \ref{gl}, we have calculated the 
free energies not only of the equilibrium state but also of the 
metastable states. 
The metastable states do not 
appear in the thermodynamic equilibrium, 
though they play important roles in actual experiments 
in which one sweeps the magnetic flux
\cite{Baelus1,Schweigert1,Kanda2,Kanda1}. 
As shown by the previous studies, 
when we change the magnetic flux the state of the system 
remains in one of the metastable states and 
it does not switches to another lower-energy state 
until the initial state finally becomes unstable. 
Although there may be effect of thermally assisted tunneling, 
such effects are limited to the very vicinity of the critical temperature. 
From these we point out a possibility that 
the {\it nodal state} may not appear at all in the field sweep experiment even though it can be a true equilibrium 
state for some values of the flux. 
This is understood from the Fig. \ref{sweep} (a), where 
the free energy for $t=0.5$ and the 
time evolution of the system under a field sweep process 
are shown. 
The bold curves show the free energies of the 
metastable states and the equilibrium state at $t=0.5$. 
The branch corresponding to the 
{\it nodal state} is indicated by a bold arrow. 
The dashed and dotted curves show the time 
evolution of the system in up- 
and down-sweep experiment, respectively, 
where it is assumed that the transition between 
different branches are prohibited 
by the energy barrier 
except when the system comes to the end of a branch. 
In this case, although the {\it nodal state} can be a 
true equilibrium state, it does not appear during the 
field sweep process. 
In contrast to thie, 
the {\it nodal state} appears during the field sweep process 
at $t=0.78$, 
as shown in Fig. \ref{sweep} (b) 
(Only up-sweep process is indicated). 
Therefore the observability of the {\it nodal state} 
in the field-sweep experiment may be 
further limited to the region in the vicinity of the 
critical temperature. 
More precise numerical simulation is required to 
clarify these dynamical processes 
of this system quantitatively, which is left for future studies.

\section{summary}
\label{summary}

In this paper, we have investigated the superconducting states 
on a M\"obius strip in terms of the numerical 
analyses based on Ginzburg-Landau and Bogoliubov-de Gennes theory. 
It has been shown that in the M\"obius geometry a novel 
{\it nodal state} can appear both in equilibrium and metastable 
states. 
The experimental observability of the {\it nodal states} is discussed 
based on the findings.

\begin{acknowledgments}

M.H and H. E were  financially supported by 
Grants-in-Aid for Scientific Research of Ministry
of Education, Science, and Culture. 
K.K was financially  supported by the Sumitomo Foundation. 
\end{acknowledgments}

%--------------------------------
\bibliographystyle{apsrev}

\vfill

\end{document}